\documentclass[nohyper,12pt,letterpaper]{JHEP3}
\usepackage{epsfig}

\def \eqn#1#2{\begin{equation}#2\label{#1}\end{equation}}

\title{de Sitter Vacua, Renormalization and Locality}

\author{T. Banks\\ SCIPP, University of California, Santa Cruz, CA 95064\\
NHETC, Rutgers University, Piscataway, NJ 08854\\

E-mail: \email banks@scipp.ucsc.edu}

\author{L. Mannelli, \\
     Department of Physics and Institute for Particle Physics\\
   University of California, Santa Cruz, CA 95064\\

E-mail: \email mannelli@physics.ucsc.edu}

\abstract{We analyze the renormalization properties of quantum field theories in de Sitter space and show that only two of the maximally invariant
vacuum states of free fields lead to consistent perturbation expansions.  One is the Euclidean vacuum and the other can be viewed as an analytic
continuation of Euclidean functional integrals on $RP^d$.  The corresponding Lorentzian manifold is the future half of global de Sitter space with
boundary conditions on fields at the origin of time.  We argue that the perturbation series in this case has divergences at the origin which
render the future evolution of the system indeterminate, without a better understanding of high energy physics.}

\begin{document}


\received{Sept. 13, 2002} \accepted{} \preprint{\hepth{0209113}\\RUNHETC-2002-34\\SCIPP-02/24}

\section{\bf Introduction}

In the recent outbreak of interest in de Sitter spacetimes, attention has been drawn again to the existence of a one (complex) parameter family of
vacuum states (called the $\alpha$-vacua) for free quantum fields in de Sitter spacetime\cite{allenmott}. Experts in the field have long harbored
a vague suspicion that only the standard Euclidean vacuum was sensible, but until now there has been no conclusive argument to this effect.  The
purpose of this note is to present one.

The argument is, in essence, very simple.  Propagators in quantum field theory are singular on the light cone.  The propagators in the
$\alpha$-vacua are linear superpositions of a Euclidean\footnote{We use the short phrase Euclidean propagator to denote the propagator of a field
in dS space, which is obtained by analytic continuation of the Euclidean functional integral on a sphere.} propagator evaluated between two points
$x,y$, and the same propagator evaluated between $x$ and the antipodal point to $y$, $y^A$. The Feynman diagrams of interacting quantum field
theory contain products of propagators between the same two points. These are not distributions, and a subtraction procedure must be supplied to
define them.  The key point of standard renormalization theory is that the subtractions all take the form of local contributions to the effective
action, and can thus be viewed as renormalizations of couplings in the theory. We will show by simple examples that in the $\alpha$-vacua this is
no longer true.  The subtractions include non-local contributions to the effective action of the form {\it e.g.} \eqn{subt}{\delta S =
\delta\lambda \int \phi (x) \phi (x^A ) } where $\delta\lambda$ is a divergent constant.   Thus, renormalized interacting field theory in a
generic $\alpha$ vacuum is intrinsically non-local, and presumably has no sensible physical interpretation.

There are only two values of $\alpha $ for which this catastrophe is avoided.  The first is $ \Re (\alpha )=a=-\infty $ which gives the standard
Euclidean vacuum and has no antipodal singularity .  The second is $\alpha = 0$, which is the unique vacuum state invariant under the antipodal
map.  The Green's function in this vacuum appears to be the analytic continuation of a Euclidean functional integral on $RP^d$\footnote{To our
knowledge, E. Witten\cite{wit} was the first to point out the significance of this special value of alpha and its Euclidean interpretation. }.  In
this vacuum state, which we call the {\it antipodal vacuum} we must view the Lorentzian spacetime manifold as the orbifold of de Sitter space by
the antipodal map.  Every point is identified with its antipode, and the interaction \ref{subt} is local. From a physical point of view we have a
manifold with a past spacelike singularity and an asymptotic de Sitter future.  We call this spacetime the {\it antipodal universe}.

In discussions of inflationary cosmology, one often invokes a Quantum No-Hair Theorem for de Sitter space .  According to this theorem, generic
initial states of quantum fields in dS space, evolve into a state indistinguishable from the Euclidean vacuum after enough e-foldings.  A crucial
assumption in this theorem, is that the initial state approaches the Euclidean vacuum for very high angular momentum modes (in global coordinates
- in planar coordinates we would say ordinary momentum modes).   Modes of any finite comoving wave number are redshifted to a size larger than the
horizon volume after a sufficient number of e-foldings, and are no longer observable by a local measurement. If the initial state is the Euclidean
vacuum for sufficiently high momentum modes, then the local observer will eventually see a state indistinguishable from the Euclidean vacuum.

The state implied by the orbifold boundary conditions does not satisfy the conditions of this theorem.  In global coordinates the Euclidean vacuum
for a boson field is a Gaussian with time dependent covariance, for each angular momentum mode.  The orbifold boundary conditions imply instead
that the initial wave function of the even angular momentum states is a field eigenstate , while that of the odd modes is an eigenstate of the
canonical momentum.   These are non-normalizable states,for each angular momentum mode, and differ from the Euclidean vacuum for arbitrarily large
angular momentum. They do not obey the de Sitter no hair theorem.  Thus, the future evolution of the antipodal universe depends on the initial
conditions.

We argue further that the initial conditions may be subject to infinite ultraviolet corrections in higher orders of perturbation theory. These are
the standard UV divergences of fixed time Schrodinger picture states in quantum field theory.  If this were true , we would have to claim that,
without a nonperturbative understanding of the state near the orbifold singularity, we could not make reliable predictions in the antipodal
universe.

These considerations cast doubt on the identification of the Lorentzian antipodal vacuum with the analytic continuation of a Euclidean functional
integral on $RP^d$.  The latter is renormalized by the standard counterterms for quantum field theory on smooth manifolds without boundary. It may
be that the boundary conditions defined by the $RP^d$ functional integral are a fixed point of the boundary renormalization group of the
Lorentzian orbifold field theory, but we have not done enough computations to verify this conjecture.

All of these arguments are made in the context of quantum field theory in a fixed spacetime background.  In quantum gravity, we have the
additional problem that the antipodal initial state has infinite energy density, which leads us to expect a large back reaction. A much more
extensive discussion of the back reaction problem in $\alpha$-vacua will be presented in \cite{KKLSS}.

Our conclusion is that only the Euclidean vacuum state has a chance of describing sensible physical processes in de Sitter space.   The rest of
this note is devoted to calculations which explicate the argument made above.

We note that after we submitted this paper to arXiv.org, two related papers
appeared which have some overlap with our work.  The first, by
Einhorn and Larsen\cite{el}, discusses aspects of higher loop graphs
in $\alpha$ vacua, and also concludes that these are generally ill-defined.
The second\cite{verl} discusses the $Z_2$ orbifold of dS space (and points
out that it was first introduced long ago by Schrodinger).  It is not clear
to us that their definition of the quantum theory is the same as ours.  They
do not discuss divergences near the origin of time in this system.

\section{\bf Interacting Scalar Field Theory in an $\alpha$ Vacuum}

In this section we will present a calculation of the two point function in a simple scalar field theory.  We hope the reader will realize that our
conclusions are quite general.  In particular, we began this project by computing the two point function of the renormalized stress tensor in an
$\alpha$ vacuum.  This computation would enter into any perturbative theory of quantum gravity in de Sitter space.  This calculation is more
divergent than any we will actually present, but exhibits the same non-locality that we find in our simple example. We decided that the extra
indices and the subtleties of covariance would only distract the reader from the main point.

\subsection{ Notation}

In the following we will consider 4-dimensional de Sitter space $ dS^{4} $. It may be realized as the manifold

\eqn{de_Sitter_manifold}{-X_{0}^{2}+X_{1}^{2}+X^{2}_{2}+X_{3}^{2}+X_{4}^{2}=l^{2}}

embedded in the 5-dimensional Minkowski space $M^{4,1}$. We will use lower case $x$ to indicate 4-dimensional coordinates on $ dS^{4}$ and upper
case $X $ to denote embedding coordinates. We will denote the antipodal points by $X^{A}\equiv - X$. Henceforth we will set $l=1 $.

We are considering an interacting scalar field theory in $ dS^{4} $ with action

\eqn{action} {S=\frac{1}{2}\int \mathrm{d}^{4}x\, (-g(x))^{\frac{1}{2}}[(\nabla \phi )^{2}-m^{2}\phi^2 -\frac{\lambda }{3!} \phi ^{3}]}

In $dS^{4} $ there is a one complex parameter, $\alpha$, family of dS invariant vacua \cite{allenmott} that we will denote $|\alpha \rangle $. The
associated de Sitter invariant family of two point Wightman functions is

\[ \langle \alpha |\phi (x)\phi (y)|\alpha \rangle =W_{\alpha }(x,y)=\]

\eqn{wight}{ n^{2}\left( W_{e}(x,y)+e^{\alpha +\alpha ^{*}}\, W_{e}(y,x)+e^{\alpha }\, W_{e}(x,y^{A})+e^{\alpha ^{*}}\, W_{e}(x^{A},y)\right) }

with

\[\alpha \in \mathcal{C}\]

\eqn{}{\Re (\alpha )=a<0 }

\eqn{alpha}{n=n(\alpha )=\frac{1}{\sqrt{1-e^{\alpha +\alpha ^{*}}}}}

Here we use the Euclidean two point Wightman function $W_{e}(x,y) $ defined in \cite{Bousso}. The Euclidean Wightman function and vacuum
correspond to $ a=-\infty $.

\subsection{Computation}

In this section we will compute a term in the \emph{1-loop} effective action,  in a general  $ \alpha $-vacuum. The computation will lead to
divergent non-local counterterms. Only the Euclidean vacuum produces a completely local counterterm action.

The 1-loop, two point contribution to the effective action in our simple field theory is

\eqn{effective_action}{ \Gamma (\phi )\sim \int \mathrm{d}^{4}x\, \mathrm{d}^{4}y\, (-g(x))^{\frac{1}{2}}(-g(y))^{\frac{1}{2}}\phi
_{cl}(x)F_{\alpha }(x,y)F_{\alpha }(x,y)\phi _{cl}(y) }

The Feynman propagator $ F_{\alpha }(x,y) $ can be expressed in terms of the Wightman functions and the parameter $ \alpha  $ as

\eqn {Feymann_GF}{ F_{\alpha }(x,y)=\Theta (x_{0}-y_{0})\, W_{\alpha }(x,y)+\Theta (y_{0}-x_{0})\, W_{\alpha }(y,x) }

\eqn {Wightman_alpha_GF}{ W_{\alpha }(x,y)=n^{2}\left( W_{e}(x,y)+e^{\alpha +\alpha ^{*}}\, W_{e}(y,x)+e^{\alpha }\, W_{e}(x,y^{A})+e^{\alpha
^{*}}\, W_{e}(x^{A},y)\right) }

with

\eqn{}{ n=n(\alpha )=\frac{1}{\sqrt{1-e^{\alpha +\alpha ^{*}}}}}

The behavior of the two point Euclidean Wightman function near the light cone is for $dS^{4}$ \eqn{Wightmann_GF_near_lc}{ W_{e}(x,y)\sim
\frac{C}{\left( x_{0}-y_{0}-i\, \epsilon \right) ^{2}-\left( x_{s}-y_{s}\right) ^{2}} ,} where $x=(x_{0},x_{s}),\, y=(y_{0},y_{s}) $ and $ C $ is
a constant whose value is not relevant for the following considerations.

We will show now that in $ F_{\alpha }^{2}(x,y) $ only the terms $W_{e}^{2} $,  having a singular behavior near the light cone of the
form

\eqn{Wight_x-y} {W_{e}^2\sim \frac{C^{2}}{(x-y)^{4}}}

\eqn{Wight_xA-y} {W_{e}^2\sim \frac{C^{2}}{(x^{A}-y)^{4}}}

\eqn{Wight_x-yA} {W_{e}^2\sim\frac{C^{2}}{(x-y^{A})^{4}}}

\eqn{Wight_x-yA_xA-y} {W_{e}^2\sim \frac{C^{2}}{(x-y^{A})^{2}(x^{A}-y)^{2}}}

contribute to the divergent part of the effective action. In these equations, we suppress the $i\epsilon $ prescription because it is not relevant
at this point. Considering $W^{2}(x,y) $ as a distribution on the space of test function $\phi (x) $ we have

\[ T_{W_{e}^{2}}[\phi ]=\int \mathrm{d}^{4}x\, W^{2}_{e}(x,y)\phi (x)\]

\[ =\int \mathrm{d}^{4}x\, \left( W^{2}_{e}(x,y)-\frac{C^{2}}{(x-y)^{4}}\right) \phi (x)+\int \mathrm{d}^{4}x\, \frac{C^{2}\phi(x)}{(x-y)^{4}}\]

\[ =\int \mathrm{d}^{4}x\, \left( W^{2}_{e}(x,y)-\frac{C^{2}}{(x-y)^{4}}\right) \phi (x)\]

\[+\int \mathrm{d}^{4}x\, \frac{C^{2}\left( \phi (x)-\phi (y)\right) }{(x-y)^{4}}+\phi (y)\int \mathrm{d}^{4}x\, \frac{C^{2}}{(x-y)^{4}}\]

\eqn{}{ =Regular+\int \mathrm{d}^{4}z\, \delta (z-y)\phi (z)\int \mathrm{d}^{4}x\, \frac{C^{2}}{(x-y)^{4}}}

where the regular part does not contribute to the divergent part of the effective action. Similarly, in the terms which contain squares of
Wightman functions evaluated between points and their antipodes, we have

\[ T_{W_{e}^{2}}[\phi ]=\int \mathrm{d}^{4}x\, W^{2}_{e}(x^{A},y)\phi (x) \]

\eqn{}{ =Regular+\int \mathrm{d}^{4}z\, \delta (z-y^{A})\phi (z)\int \mathrm{d}^{4}x\, \frac{C^{2}}{(x^{A}-y)^{4}}}

and

\[ T_{W_{e}^{2}}[\phi ]=\int \mathrm{d}^{4}x\, W^{2}_{e}(x,y^{A})\phi (x) \]

\eqn{}{ =Regular+\int \mathrm{d}^{4}z\, \delta (z-y^{A})\phi (z)\int \mathrm{d}^{4}x\, \frac{C^{2}}{(x-y^{A})^{4}}}

Similarly

\[ T_{W_{e}^{2}}[\phi ]=\int \mathrm{d}^{4}x\, W_{e}(x^{A},y)W_{e}(x,y^{A})\phi (x)\]

\eqn{}{ =Regular+\int \mathrm{d}^{4}z\, \delta (z-y^{A})\phi (z)\int \mathrm{d}^{4}x\, \frac{C^{2}}{(x^{A}-y)^{2}(x-y^{A})^{2}}}

All the other terms in $ W_{e}^{2}$ are regular and do not contribute to the divergent part of the effective action.

After eliminating the regular terms in $F_{\alpha }^{2}(x,y) $  and doing the replacements $\Theta (x_{0}-y_{0})\, \Theta (y_{0}-x_{0})\rightarrow
0$, $ \Theta (x_{0}-y_{0})^{2}\rightarrow \Theta (x_{0}-y_{0}) $, and $ \Theta (y_{0}-x_{0})^{2}\rightarrow \Theta (y_{0}-x_{0}),$

we get

\[ F_{\alpha }(x,y)^{2}=n^{4}\, \Theta (x_{0}-y_{0})\, W_{e}(x,y)^{2}+e^{2\, \alpha +2\, \alpha ^{*}}\, n^{4}\, \Theta (y_{0}-x_{0})\,
W_{e}(x,y)^{2}\]

\[ +e^{2\, \alpha ^{*}}\, n^{4}\, \Theta (x_{0}-y_{0})\, W_{e}(x,y^{A})^{2}+2\, e^{\alpha +\alpha ^{*}}\, n^{4}\, \Theta (x_{0}-y_{0})\,
W_{e}(x,y^{A})\, W_{e}(x^{A},y)\]

\[ +e^{2\, \alpha }\, n^{4}\, \Theta (x_{0}-y_{0})\, W_{e}(x^{A},y)^{2}+2\, e^{\alpha +\alpha ^{*}}\, n^{4}\, \Theta (x_{0}-y_{0})\, W_{e}(x,y)\,
W_{e}(y,x)\]

\[ +2\, e^{\alpha +\alpha ^{*}}\, n^{4}\, \Theta (y_{0}-x_{0})\, W_{e}(x,y)\, W_{e}(y,x)+e^{2\, \alpha +2\, \alpha ^{*}}\, n^{4}\, \Theta
(x_{0}-y_{0})\, W_{e}(y,x)^{2}\]

\[ +n^{4}\, \Theta (y_{0}-x_{0})\, W_{e}(y,x)^{2}+e^{2\, \alpha ^{*}}\, n^{4}\, \Theta (y_{0}-x_{0})\, W_{e}(y,x^{A})^{2}\]

\eqn{}{ +2\, e^{\alpha +\alpha ^{*}}\, n^{4}\, \Theta (y_{0}-x_{0})\, W_{e}(y,x^{A})\, W_{e}(y^{A},x)+e^{2\, \alpha }\, n^{4}\, \Theta
(y_{0}-x_{0})\, W_{e}(y^{A},x)^{2}}

Replacing the $W_{e}$ terms with their singular behavior near the light cone,  we find

\[ F_{\alpha }(x,y)^{2}\sim \]

\[\delta (x-y)\, \left( \frac{C^{2}\, n^{4}\, \Theta (x_{0}-y_{0})}{\left( \left( x_{0}-y_{0}-i\, \epsilon \right) ^{2}-\left(
x_{s}-y_{s}\right) ^{2}\right) ^{2} }+\frac{C^{2}\, e^{2\, \alpha +2\, \alpha ^{*}}\, n^{4}\, \Theta (x_{0}-y_{0})}{\left( \left( y_{0}-x_{0}-i\,
\epsilon \right) ^{2}-\left( x_{s}-y_{s}\right) ^{2}\right) ^{2}} \right. \]

\[+\frac{2\, C^{2}\, e^{\alpha +\alpha ^{*}}\, n^{4}\, \Theta (x_{0}-y_{0})}{\left( \left( x_{0}-y_{0}-i\, \epsilon \right) ^{2}-\left(
x_{s}-y_{s}\right) ^{2}\right) \, \left( \left( y_{0}-x_{0}-i\, \epsilon \right) ^{2}-\left( x_{s}-y_{s}\right) ^{2}\right) }\]

\[ +\frac{C^{2}\, e^{2\, \alpha +2\, \alpha ^{*}}\, n^{4}\, \Theta (y_{0}-x_{0})}{\left( \left( x_{0}-y_{0}-i\, \epsilon \right) ^{2}-\left(
x_{s}-y_{s}\right) ^{2}\right) ^{2}}+\frac{C^{2}\, n^{4}\, \Theta (y_{0}-x_{0})}{\left( \left( y_{0}-x_{0}-i\, \epsilon \right) ^{2}-\left(
x_{s}-y_{s}\right) ^{2}\right) ^{2}}\]

\[ \left. +\frac{2\, C^{2}\, e^{\alpha +\alpha ^{*}}\, n^{4}\, \Theta (y_{0}-x_{0})}{\left( \left( x_{0}-y_{0}-i\, \epsilon \right) ^{2}-\left(
x_{s}-y_{s}\right) ^{2}\right) \, \left( \left( y_{0}-x_{0}-i\, \epsilon \right) ^{2}-\left( x_{s}-y_{s}\right) ^{2}\right) } \right) \]

\[ +\delta (x-y^{A})\, \left( \frac{C^{2}\, e^{2\, \alpha }\, n^{4}\, \Theta (x_{0}-y_{0})}{\left( \left( x_{0}^{A}-y_{0}-i\, \epsilon \right)
^{2}-\left( x_{s}^{A}-y_{s}\right) ^{2}\right) ^{2}} +\frac{C^{2}\, e^{2\, \alpha ^{*}}\, n^{4}\, \Theta (x_{0}-y_{0})}{\left( \left(
x_{0}-y_{0}^{A}-i\, \epsilon \right) ^{2}-\left( x_{s}-y_{s}^{A}\right) ^{2}\right) ^{2}} \right. \]

\[ +\frac{2\, C^{2}\, e^{\alpha +\alpha ^{*}}\, n^{4}\, \Theta (x_{0}-y_{0})}{\left( \left( x_{0}^{A}-y_{0}-i\, \epsilon \right) ^{2}-\left(
x_{s}^{A}-y_{s}\right) ^{2}\right) \, \left( \left( x_{0}-y_{0}^{A}-i\, \epsilon \right) ^{2}-\left( x_{s}-y_{s}^{A}\right) ^{2}\right) }\]

\[ +\frac{C^{2}\, e^{2\, \alpha ^{*}}\, n^{4}\, \Theta (y_{0}-x_{0})}{\left( \left( y_{0}-x_{0}^{A}-i\, \epsilon \right) ^{2}-\left(
x_{s}^{A}-y_{s}\right) ^{2}\right) ^{2}}+\frac{C^{2}\, e^{2\, \alpha }\, n^{4}\, \Theta (y_{0}-x_{0})}{\left( \left( y_{0}^{A}-x_{0}-i\, \epsilon
\right) ^{2}-\left( x_{s}-y_{s}^{A}\right) ^{2}\right) ^{2}}\]

\eqn{}{ \left. +\frac{2\, C^{2}\, e^{\alpha +\alpha ^{*}}\, n^{4}\, \Theta (y_{0}-x_{0})}{\left( \left( y_{0}-x_{0}^{A}-i\, \epsilon \right)
^{2}-\left( x_{s}^{A}-y_{s}\right) ^{2}\right) \, \left( \left( y_{0}^{A}-x_{0}-i\, \epsilon \right) ^{2}-\left( x_{s}-y_{s}^{A}\right)
^{2}\right) }\right) }

The \( \delta (x-y^{A}) \) term gives rise to a non local, divergent, contribution to the effective action . The coefficient of \( \delta
(x-y^{A}) \) is

\[ \left( \frac{C^{2}\, e^{2\, \alpha }\, n^{4}\, \Theta (x_{0}-y_{0})}{\left( \left( x_{0}^{A}-y_{0}-i\, \epsilon \right) ^{2}-\left(
x_{s}^{A}-y_{s}\right) ^{2}\right) ^{2}}+\frac{C^{2}\, e^{2\, \alpha ^{*}}\, n^{4}\, \Theta (x_{0}-y_{0})}{\left( \left( x_{0}-y_{0}^{A}-i\,
\epsilon \right) ^{2}-\left( x_{s}-y_{s}^{A}\right) ^{2}\right) ^{2}}\right. \]

\[ +\frac{2\, C^{2}\, e^{\alpha +\alpha ^{*}}\, n^{4}\, \Theta (x_{0}-y_{0})}{\left( \left( x_{0}^{A}-y_{0}-i\, \epsilon \right) ^{2}-\left(
x_{s}^{A}-y_{s}\right) ^{2}\right) \, \left( \left( x_{0}-y_{0}^{A}-i\, \epsilon \right) ^{2}-\left( x_{s}-y_{s}^{A}\right) ^{2}\right) }\]

\[ +\frac{C^{2}\, e^{2\, \alpha ^{*}}\, n^{4}\, \Theta (y_{0}-x_{0})}{\left( \left( y_{0}-x_{0}^{A}-i\, \epsilon \right) ^{2}-\left(
x_{s}^{A}-y_{s}\right) ^{2}\right) ^{2}}+\frac{C^{2}\, e^{2\, \alpha }\, n^{4}\, \Theta (y_{0}-x_{0})}{\left( \left( y_{0}^{A}-x_{0}-i\, \epsilon
\right) ^{2}-\left( x_{s}-y_{s}^{A}\right) ^{2}\right) ^{2}}\]

\eqn{}{ \left. +\frac{2\, C^{2}\, e^{\alpha +\alpha ^{*}}\, n^{4}\, \Theta (y_{0}-x_{0})}{\left( \left( y_{0}-x_{0}^{A}-i\, \epsilon \right)
^{2}-\left( x_{s}^{A}-y_{s}\right) ^{2}\right) \, \left( \left( y_{0}^{A}-x_{0}-i\, \epsilon \right) ^{2}-\left( x_{s}-y_{s}^{A}\right)
^{2}\right) }\right)}

After the substitutions $(x_{0}^{A},\, x_{s}^{A})\rightarrow (-x_{0},\, -x_{s})$, $(y_{0}^{A},\, y_{s}^{A})\rightarrow (-y_{0},\, -y_{s})$, $
\Theta (x_{0}-y_{0})+\Theta (y_{0}-x_{0})\rightarrow 1$,

we find that the non-local part of the divergent counterterm is

\[\left( \frac{C^{2}\, e^{2\, \alpha }\, n^{4}}{\left( \left( x_{0}+y_{0}+i\, \epsilon \right) ^{2}-\left( x_{s}+y_{s}\right) ^{2}\right)
^{2}}+\frac{C^{2}\, e^{2\, \alpha ^{*}}\, n^{4}}{\left( \left( x_{0}+y_{0}-i\, \epsilon \right) ^{2}-\left( x_{s}+y_{s}\right) ^{2}\right)
^{2}}\right. \]

\eqn{non_local_part} {\left. +\frac{2\, C^{2}\, e^{\alpha +\alpha ^{*}}\, n^{4}}{\left( \left( x_{0}+y_{0}+i\, \epsilon \right) ^{2}-\left(
x_{s}+y_{s}\right) ^{2}\right) \, \left( \left( x_{0}+y_{0}-i\, \epsilon \right) ^{2}-\left( x_{s}+y_{s}\right) ^{2}\right) }\right) }

The three terms in this expression are different because they have distinct $ i\epsilon  $ prescriptions and therefore diverse poles in the
complex plane.

As a consequence to eliminate all the divergent, non local terms in the effective action we must set

\eqn{}{ e^{\alpha }=e^{a+ib}=0}

\eqn{}{ e^{\alpha ^{*}}=e^{a-ib}=0}

\eqn{}{ \Rightarrow e^{a}=0}

\eqn{}{ \Rightarrow a=-\infty }

This corresponds to the choice of the Euclidean vacuum as previously stated. We should remark that the constant $ n=\frac{1}{\sqrt{1-e^{\alpha
+\alpha ^{*}}}}=\frac{1}{\sqrt{1-e^{2a}}}$. can never be zero because the family of de Sitter invariant vacua is defined by $ \alpha \in
\mathcal{C}$, $ \Re (\alpha )=a<0$.  There is however another way to obtain a system with local effective action.  The nonlocalities are all
products of fields at points and their antipodes.  For $\alpha = 0$ we can interpret the Green's functions as living on an orbifold of dS space,
the antipodal universe, in which a point is identified with its antipode. On this spacetime, all of our counterterms can be viewed as local
operators.

Witten\cite{wit} has suggested that for this value of $\alpha$ the Green's functions can be viewed as analytic continuations of the Euclidean
functional integral on the real projective space $RP^4$.   Since $RP^4$ is a smooth manifold without boundary, this Euclidean functional integral
should be renormalized by the same local counterterms that define the field theory on the sphere.  We will discuss this interpretation in the next
section.

\section{\bf The Wave Functional in the Antipodal Vacuum}

We have seen that, with the exception of the Euclidean and Antipodal vacua, field theory in an $\alpha$ vacuum cannot be renormalized by local
counterterms.   We now want to investigate whether the Antipodal vacuum forms the basis for a sensible quantum field theory.  Certainly, the
Euclidean functional integral on $RP^d$ is well defined.  However, it is not immediately apparent that the Green's functions defined by this
functional integral have a Hamiltonian interpretation.  The conventional reflection positivity argument requires the reflected Euclidean points to
be distinct from the points themselves.

Indeed, it would appear that the Lorentzian version of the Antipodal universe requires more renormalization than the corresponding Euclidean
functional integral.  $RP^d$ is a smooth manifold without boundary and the Euclidean functional integral on this manifold will be renormalized by
the same local subtractions that are required for the Euclidean functional integral on the sphere. However, the Lorentzian version of the theory
describes the evolution of a quantum field theory starting from a fixed state at a sharp time.  It has been known since the work of
Symanzik\cite{sym} that in renormalizable quantum field theories, the wave functional at a sharp time requires additional renormalizations, above
and beyond those which render the Green's functions finite.  In modern parlance, the sharp time state introduces a boundary into the system and
one must introduce counterterms for all relevant boundary operators at the fixed point of the bulk renormalization group.

Thus, it would seem that, if field theory is to be defined in the antipodal vacuum it requires additional definitions to determine the initial
state.   These remarks also seem to indicate that the connection between the Lorentzian theory and the Euclidean theory on $RP^d$ must somehow be
valid only in the absence of boundary renormalizations.  We have remarked that the Euclidean antipodal Green's functions do not seem to require
additional subtractions. It is possible that this means that the Lorentzian boundary conditions implied by continuation from $RP^d$ are
automatically fixed points of the boundary renormalization group.

Indeed, the above discussion of boundary renormalization is valid for boundary conditions of the form
$\phi (t=0 , x) = \phi_0 (x)$, which would define the Schrodinger wave functional.  We then think of
the orbifold boundary condition as a restriction on the allowed Schrodinger functionals.  Perhaps , since the Lorentzian orbifold does not have a geometric boundary, all boundary counterterms will vanish in such
a state\footnote{TB thanks M. Douglas for a discussion of this point.}.
We have not been able to determine the validity of such a conjecture. In
particular, in general field theories there would seem to be marginal and relevant boundary operators which are not projected out by the orbifold
condition.  We do not understand why additional counterterms proportional to these relevant operators are not generated by the Lorentzian Feynman
rules.

Even apart from these additional renormalization effects, the state defined by the antipodal boundary conditions is somewhat singular.
Classically, the field is required to be invariant under simultaneous reflection in the global coordinate time and spatial sphere.  If we expand
the field into spherical harmonics, then (for free field theory in dS space) each mode $\phi_L$ is a time dependent harmonic oscillator, with a
frequency that is even under reflection about the point of minimal size.   Under reflection in the sphere, $\phi_L \rightarrow (- 1)^L \phi_L$.
Thus, invariance under the antipodal map is equivalent to the quantum mechanical statement that the initial state is annihilated by $\phi_L$ for
odd $L$ and by the conjugate momentum $\Pi_L$ for even $L$.  The quantum system is then studied as a collection of time dependent oscillators,
with these initial conditions, on the interval $t \in [0,\infty ]$.   Note that the quantum state defined by this boundary condition differs from
the Euclidean state of the same system even for $L\rightarrow\infty$.  The dS No-Hair Theorem is not applicable, and this state does not approach
the Euclidean vacuum at large times.

Finally, we note that the deviation from the Euclidean vacuum for large $L$ also implies that the matrix elements of the renormalized stress
tensor between states of the form

\eqn{states}{a^{\dagger}_{L_1} \ldots a^{\dagger}_{L_n} |A>}

where $|A>$ is the antipodal vacuum and the operators are its associated creation operators, will blow up as $t \rightarrow 0$.

These additional divergences have little to do with renormalization.  They are more analogous to the singularities at particular places in
Lorentzian momentum space that one finds in the analytic continuation of renormalized Euclidean Green's functions in any field theory.  That is,
they represent real physical processes, rather than virtual contributions to the effective action.

The consequence of these remarks is that, although the antipodal vacuum does not suffer from the
renormalization problems of the generic $\alpha$ vacuum, its physics is not under control at $t=0$.

\section{Conclusion}

We have investigated the perturbative renormalizability of quantum field theories in rigid dS space, when the vacuum state of the free fields is
chosen to be one of the non-Euclidean, dS invariant vacua.  In general the renormalization program fails.  Non-local counterterms, involving
products of fields at both points and their antipodes, are necessary to render the interacting Green's functions finite.  Even if it were possible
to prove that this nonlocal renormalization program could be carried out to all orders (which is by no means obvious) , the resulting theory would
probably not have a Hamiltonian interpretation. We consider this as evidence that quantum field theory in generic $\alpha$-vacua does not make
sense.

Apart from the Euclidean vacuum, the antipodal vacuum is the only one where the non-local renormalization problem can be avoided. This vacuum
state can be interpreted in terms of field theory on an orbifold of dS space, in which the non-local operators are local.  It is possible that the
resulting theory is just the analytic continuation of the Euclidean functional integral on $RP^d$, though one would have to do a more thorough
study of boundary renormalizations in the Lorentzian orbifold in order to prove this.

Independently of this renormalization problem, there are clearly divergent matrix elements of local operators like the stress tensor on the fixed
plane $ t = 0$ of the Lorentzian orbifold. If we tried to couple gravity to the system this would lead to large back reaction effects.  At the
very least, a straightforward perturbative approach to the system would fail.  Back reaction effects and the failure of the semiclassical
approximation in general $\alpha$-vacua are discussed in more detail in \cite{KKLSS}.
\section{Acknowledgments}

The research of T.B and L.M. was supported in part by DOE grant number DE-FG03-92ER40689.  We would like to thank S.Shenker and L. Susskind for
discussing their work with us prior to publication.  T.B. also acknowledges informative discussions with
M. Douglas and E. Witten

\end{document}